\documentclass[12pt]{article}
\usepackage{amsmath,amssymb,amsfonts} 
\usepackage{graphics}                 
\usepackage{color}                    
\usepackage{hyperref}                 
\usepackage[cp1252]{inputenc}         
\usepackage[T1]{fontenc}
\usepackage{epsfig}

\usepackage{authblk}

\title{Impact of DM direct searches and the LHC analyses on branon phenomenology}

\author[1]{Jose A. R. Cembranos}
\author[2]{J. Lorenzo D\'{\i}az-Cruz}
\author[2]{Lilian Prado}
\affil[1]{Departamento de F\'{\i}sica Te\'orica I, Universidad
Complutense de Madrid, E-28040, Spain} \affil[2]{Facultad de
Ciencias F\'{\i}sico-Matem\'aticas, Benem\'erita Universidad
Aut\'onoma de Puebla, C.P. 72570, Puebla, Pue., Mexico}

\date{\today}

\begin{document}

\maketitle
\begin{abstract}

Dark Matter direct detection experiments are able to exclude
interesting parameter space regions of particle models which predict
an important amount of thermal relics. We use recent data to
constrain the branon model and to compute the region that is favored
by CDMS measurements. Within this work, we also update present colliders
constraints with new studies coming from the LHC. Despite the present
low luminosity, it is remarkable that for heavy branons, CMS and
ATLAS measurements are already more constraining than previous
analyses performed with TEVATRON and LEP data.
\end{abstract}

pacs 95.35.+d, 11.10.Kk, 11.25.Wx

\section{Introduction}

Identifying the nature of Dark Matter (DM) is a central problem in
contemporary physics. The nature of DM and how it fits into our
current understanding of elementary particles, is not known yet.
Numerous astrophysical and cosmological data require the existence
of a DM component, that accounts for about 20\% of the energy
content of our universe. Although there are other possibilities
\cite{other}, DM is usually assumed to be in the form of stable
Weakly-Interacting Massive Particles (WIMPs) that naturally
freeze-out with the right thermal abundance, with a mass of the
order of the electro-weak (EW) scale. The most studied
 DM candidate is the neutralino, which can be
identified as the lightest supersymmetric particle in many
supersymmetric models. These models own a rich phenomenology due
to the large number of new particles present in the theory \cite{kamion}.

Besides, the existence of extra dimensions are well motivated
theoretically and DM candidates arise associated to new degrees of
freedom. In particular, branons have been proposed to explain the
missing matter problem within the so-called brane-world scenario
(BWS) \cite{BW1}. In this framework, the SM fields are forced to
live on a three-dimensional hypersurface called brane, while gravity
propagates on the higher $D=4+N$ dimensional bulk space. The
fundamental scale of gravity is not the Planck scale $M_{p}$
anymore, but a new scale $M_{D}$ which in this work will be
considered arbitrary \cite{BW1}.

In this scenario, the existence of extra dimensions generates new
fields on the brane, giving rise to a Kaluza-Klein (KK) tower of
massive gravitons. The brane has a finite tension $f^{4}$ and its
fluctuations are parametrized by the so-called branon fields
$\pi^{a}$. These branons are the massless Goldstone bosons arising
from the spontaneous breaking of the exact symmetry existing in the
case of translational invariance in the bulk space. In the general
case of translational invariance explicitly broken, the branons are
expected to be massive fields. When these branons are considered,
the coupling of the Standard Model (SM) particles to any bulk field
is exponentially suppressed by a factor
$exp[-M^{2}_{KK}\Lambda^{2}/8\pi^{}f^{4}]$, where $M_{KK}$ is the mass
of the corresponding KK mode of the tower of massive gravitons and
$\Lambda$ is the cut-off of the effective theory that describes the branon
phenomenology \cite{bando,BWRad}.

If the tension scale $f$ is much smaller than the other 
new scales so that $f^2\ll \Lambda\,M_{KK}$, then the KK modes
decouple from the SM particles, so that at low energies the only
brane-world related particles that must be taken into account are branons.
In the following we will assume this to be the case and accordingly we will
deal only with SM particles and branons \cite{BW1}.

$\Lambda$ could represent the width of brane or any other mechanism 
that modified the short-distance theory to cure the ultraviolet behavior 
of branons. However, for our purposes, $\Lambda$ is just a
phenomenological parameter. From the point of view of the effective theory,
$\Lambda$/f parameterizes how strongly (or weakly) coupled quantum brane is, and
therefore controls the unknown relative importance of tree-level versus loop
branon effects. In \cite{BWRad} it was shown that the perturbative loop analysis only
makes sense for approximately $\Lambda\lesssim 4\,\pi^{1/2}f\,N^{-1/4}$. We
will work in such a limit.

These branons are expected to be nearly massless and weakly
interacting at low energies. In general, translational invariance in
the extra dimensions is not necessarily an exact symmetry, so that
explicit symmetry breaking leads to a branon mass $M$. Brane
fluctuations could be candidates for the cosmological DM and they
could also make up the galactic halo and explain the local dynamics.
Therefore, they could be detected by DM search experiments
\cite{BW1, BW2}.

Several experiments have been developed to detect DM directly and
indirectly. The direct DM detection experiments are designed to
observe the elastic scattering of DM particles with nuclei while
indirect DM searches may detect the DM annihilation productions such
as protons, antiprotons, electrons, positrons, neutrinos and gamma
rays. Complementary to these, collider DM searches are performed at
experiments like CERN (European Organization for Nuclear Research)
LHC (Large Hadron Collider) \cite{Coll, BWHad}.

The outline of the paper goes as follows: first we shall discuss briefly
the model in Section~\ref{FBW}.
Section~\ref{direct} is dedicated to direct DM search experiments,
reporting some DM candidate events and their relation with branons.
Section~\ref{collider} is devoted to the analysis of branons related
to collider results and new LHC constraints.
Finally section~\ref{conclusion} presents our conclusions.

\section{Flexible Brane-Worlds}\label{FBW}

The presence of branes in extra dimensional models have been studied
from many different points of view (read, for instance, \cite{Klebanov:1997kc}).
We will consider a single-brane model in large extra dimensions, where
the four-dimensional space-time $M_{4}$ is embedded in a
D-dimensional bulk space. This space is assumed to have the form
$M_{D}=M_{4} \times B$ where the B homogeneous space is an
N-dimensional compact manifold, such that $D=4+N$. The brane lies
along $M_{4}$ and its contribution to the bulk gravitational field
is neglected \cite{BW1,BW2,SKCP}.

Branons couple to the conserved energy-momentum tensor of the
Standard Model evaluated in the background metric $T^{\mu\nu}_{SM}$.
The lowest order effective Lagrangian is given by \cite{BW1}:

\begin{eqnarray}
\mathcal{L}_{Br} &=&\frac{1}{2}g^{\mu \upsilon }\partial _{\mu }\pi
^{\alpha }\partial _{\nu }\pi ^{\alpha }-\frac{1}{2}M^{2}\pi
^{\alpha }\pi ^{\alpha }  \nonumber
\\
&&+\frac{1}{8f^{4}}\left( 4\partial _{\mu }\pi ^{\alpha }\partial
_{\nu }\pi ^{\alpha }-M^{2}\pi ^{\alpha }\pi ^{\alpha }g_{\mu
\upsilon }\right) T_{SM}^{\mu \upsilon }. \label{lag}
\end{eqnarray}

Branons are stable and difficult to detect because they always
interact by pairs and their interactions are suppressed by the
tension scale $f$, which implies that they can be weakly
interacting. Also they are expected to be massive so that their
freeze-out temperature could be relatively high and then their relic
abundances could be cosmologically important. This implies that
branons are natural candidates to DM in a scenario where $f \ll
M_{D}$ \cite{BW1}.

The thermal relic branon abundance is calculated in \cite{BW1}
where it has been considered the relativistic (hot) and
nonrelativistic (cold) cases at decoupling. The allowed region for
hot branons masses are much smaller than those in neutrino dark
matter models so that they decouple much earlier than neutrinos;
therefore, hot branons are disfavored. Branons could be responsible
for the observed cosmological DM density provided $\Omega_{Br}
h^{2}=0.110\pm0.006$ \cite{pdg}.

\section{DM Direct detection experiments}\label{direct}
Several DM direct detection experiments have presented recently some
results. Their reported limits are compatible with the branon
scenario.

Assuming that the DM halo of the Milky Way is composed of branons,
its flux on the Earth is of order $10^{5}(100$ GeV/M $)$ cm$^{-2}$s$^{-1}$,
and could be sufficiently large to be measured in direct detection
experiments such as DAMA, XENON100, CoGeNT, CDMS II or
EDELWEISS-II. These experiments measure the rate $R$, and energies
$E_{R}$ of the nuclear recoils \cite{cerdeno}.

The differential counting rate for a nucleus with mass $m_{N}$ is
\begin{equation}
\frac{dR}{dE_{d}}=\frac{\rho _{0}}{m_{N} M}\int_{v_{\min
}}^{\infty }vf\left( v\right) \frac{d\sigma _{BrN}}{dE_{R}}\left(
v,E_{R}\right) dv,
\end{equation}
where $\rho_{0}$ is the local branon density, $(d\sigma
_{BrN}/dE_{R})(v,E_{R})$ is the differential cross-section for the
Branon-nucleus elastic scattering and $f(v)$ is the Branon speed
distribution in the detector frame normalized to unity.

The relative speed of a dark matter particle is of order 100
$km^{-1}s^{-1}$, so the elastic scattering is non-relativistic. Then
the recoil energy of the nucleon in terms of the scattering angle in
the center of mass frame $\theta^{*}$, and the branon-nucleus
reduced mass $\mu_{N}=Mm_{N}/(M+m_{N})$, is given by \cite{cerdeno}
\begin{equation}
E_{R}=\frac{\mu _{N}^{2}v^{2}\left( 1-\cos \theta ^{\ast }\right)
}{m_{N}}.
\end{equation}

The lower limit of the integration over the WIMP speed is given in
terms of the minimum branon speed which can cause a recoil of energy
$E_{R}$ and is given by $v_{min}=(m_{N}E_{R}/2\mu_{N}^{2})^{1/2}$.
The upper limit is infinite; however the local escape speed
$v_{esc}$ is the maximum speed in the Galactic rest frame for WIMPs
which are gravitationally bound to the Milky Way. The standard value
for this scape velocity is $v_{esc}=650$ $km s^{-1}$ \cite{cerdeno}.

Integrating the differential event rate over all the possible recoil
energies, it is possible to find the total event rate of branon
collisions with matter per kilogram per day, $R$. The smallest
recoil energy that the detector is capable of measuring is called
threshold energy, $E_{T}$. In terms of this threshold energy,
$E_{T}$, the total event rate $R$ has the form \cite{cerdeno}:
\begin{equation}
R=\int_{E_{T}}^{\infty }dE_{R}\frac{\rho _{0}}{m_{N}M}
\int_{v_{\min }}^{\infty }vf\left( v\right) \frac{d\sigma
_{Br N}}{dE_{R}}\left( v,E_{R}\right) dv.
\end{equation}

The branon-nucleus differential cross section contains the particle
physics inputs and depends on the branon-quark interaction given by
(\ref{lag}). For a general DM candidate, its nucleus cross section
is separated into a spin-independent (SI) scalar contribution and a
spin-dependent (SD) one \cite{cerdeno}:
\begin{equation}
\frac{d\sigma _{N}}{dE_{R}}=\frac{m_{N}}{2\mu _{N}^{2}v^{2}}\left(
\sigma _{0}^{SI}F_{SI}^{2}\left( E_{R}\right) +\sigma
_{0}^{SD}F_{SD}^{2}\left( E_{R}\right) \right).
\end{equation}
Here the form factors $F_{SI}(E_{R})$ and $F_{SD}(E_{R})$ account
for the coherence loss which leads to a suppression in the event
rate for heavy WIMPs on nucleons and includes the dependence on the
momentum transfer $q=\sqrt{2m_{N}E_{R}}$. $\sigma_{0}^{SI}$ and
$\sigma_{0}^{SD}$ are the spin-independent and spin-dependent cross
sections respectively, at zero momentum transfer. These quantities
still depend on nuclear structure through isospin content; that is,
the number of protons vs.\ neutrons \cite{kopp}.

In the branon case, the entire interaction is SI and can be written,
in general, as \cite{kopp}

\begin{equation}
\sigma^{\rm SI}=\frac{[Zf_p+(A-Z)f_n]^2}{f_p^2}\frac{\mu_{DM
n}^2}{\mu_{DM p}^2}\sigma_p^{\rm SI},
\end{equation}
with A the atomic mass number, Z the charge of the nucleus,
$f_{p,n}$ the SI DM couplings to proton and neutron respectively.
$\mu_{DM p}$ is the reduced DM--proton mass, and $\sigma_p^{SI}$ the
SI cross section for scattering of DM on a proton. In particular,
branons do not violate isospin symmetry if we neglect the difference
in mass of protons and neutrons. Therefore, within this
approximation: $f_{p}=f_{n}$, and $\mu\equiv\mu_{Br n}=\mu_{Br p}$.
Indeed, the branon-nucleon cross section $\sigma_{n}$ can be written
as \cite{BW1}
\begin{equation}
\sigma_p^{\rm SI}=\sigma _{n}=\frac{9M^{2}{m_n}^{2}\mu ^{2}}{64\pi f^{8}}\,,
\end{equation}
where $m_n$ is the nucleon mass. Recently, direct search experiments
have reported possible candidate events for DM. The annual modulation
signature found either by former DAMA/NAI and DAMA/LIBRA detector
localized in Gran Sasso National Laboratory points out to a light WIMP
\cite{dama}. A similar conclusion can be obtained from the CoGeNT observations.
This ultralow-noise germanium detector operated deep underground in
Soudan Underground Laboratory has found some events consistent with
a WIMP of mass 7 to 11 GeV \cite{CoGeNT}. However, these
measurements are in clear tension with exclusion limits obtained by
other experiments located in the same laboratories such as XENON100
(a liquid xenon detector at the Gran Sasso National Laboratory)
\cite{xenon}, or CDMS II (a germanium and silicon detector at the
the Soudan Underground Laboratory) \cite{CDMSevents}.

The CDMS Collaboration have reported two candidate events for DM
that are consistent with heavier WIMPs and with the present
constraints \cite{CDMSevents}. Indeed, the analysis was performed on
data taken during four periods between July 2007 and September 2008,
maximizing the expected sensitivity for a 60 GeV WIMP. Their
spectrum-averaged equivalent exposure for a WIMP of this mass is
194.1 kg-days. Two events in the WIMP acceptance region were
observed at recoil energies of 12.3 keV and 15.5 keV.

Neutrons with energies of several MeV can generate single-scatter
nuclear recoils that are indistinguishable from possible DM
interactions. Also an approximation made during the ionization pulse
reconstruction degrades the timing-cut rejection for a small
fraction of surface events with ionization energy below 6 keV.
Therefore, the probability to have observed two or more surface
events in this exposure is estimated to 20\%. Including the neutron
background, the candidate events have a probability 23\% of not
being due to a DM signal \cite{CDMSevents}.

\begin{figure}
\begin{center}\includegraphics[width=12cm]{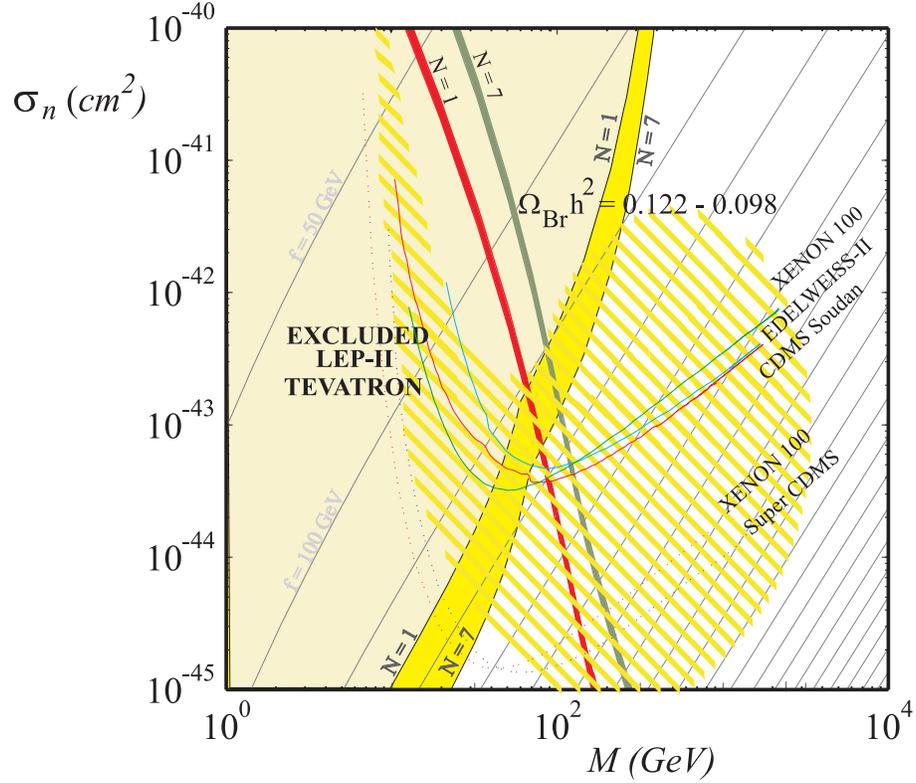} \end{center}
\caption{Elastic branon-nucleon cross section $\sigma_{n}$ in terms
of the branon mass. The two thick lines correspond to the
$\Omega_{Br}h^{2}=0.122 - 0.098$ curve for cold branons with $N=1$
(left) and $N=7$ (right). The shaded areas on the left are the
collider exclusion regions \cite{BW2,BWHad}, also for N=1, 7.
The solid lines correspond to the current limits on the
spin-independent cross section from direct detection experiments
XENON 100 \cite{xenon}, EDELWEISS-II \cite{edelweiss} and CDMS \cite{CDMSevents}. The
striped area is favored by CDMS measurements \cite{CDMSevents} and
the dotted lines are sensitivity prospects for XENON 100 and
super CDMS.} \label{figJose}
\end{figure}

The results of the analysis for branon-nucleon cross section
$\sigma_{n}$ in terms of the branon mass and the limits of direct DM
search experiments are shown in Figure \ref{figJose}. For reference,
lines of constant $f$ with 50 GeV separation are shown. The area on
the left of the $\Omega_{Br}h^{2}=0.122 - 0.098$ curves is excluded
by branon overproduction, but the right portion is compatible with
observations. Such region corresponds to $f\gtrsim 120$ GeV and
$M\gtrsim 40$ GeV. The dashed yellow region is favoured by the 2
CDMS II events at 12.3 keV and 15.5 keV at the 90 \% confidence
level.

\section{Colliders Searches at the LHC}\label{collider}
If the new physics able to explain the DM puzzle is related to the
electroweak scale, it may be accessible at the LHC and in the new
generation of collider experiments \cite{Coll,BWHad}.

\begin{table}[bt]
\centering \small{
\begin{tabular}{||c|cccc||}
\hline Experiment
&
$\sqrt{s}$(TeV)& ${\mathcal
L}$(pb$^{-1}$)&$f_0$(GeV)&$M_0$(GeV)\\
\hline
%
%
HERA$^{\,1}
$& 0.3 & 110 &  16 & 152
\\
Tevatron-II$^{\,1}
$& 2.0 & $10^3$ &  256 & 902
\\
Tevatron-II$^{\,2}
$& 2.0 & $10^3$ &   240 & 952
\\
LEP-II$^{\,2}
$& 0.2 & 600 &  180 & 103
\\
\hline
LHC$^{\,1}
$& 7 & $3.4\times10^{-4}$ &  189 & 3240
\\
LHC$^{\,1} $& 7 & $11.7\times10^{-3}$ &  236 & 3240
\\
LHC$^{\,2}
$& 7 & $1.21\times10^{-3}$ &   152 & 3390
\\

\hline
LHC$^{\,1}
$& 14 & $10^5$ &  1075 & 6481
\\
LHC$^{\,2}
$& 14 & $10^5$ &   797 & 6781
\\
\hline
\end{tabular}
} \caption{\footnotesize{Limits from direct branon searches at
colliders (results at the $95\;\%$ c.l.). Upper indices $^{1,2}$
denote monojet and single photon channels respectively. Current data
\cite{adriani, weng, schw} and prospects for the LHC are compared
with present constraints from LEP \cite{Coll}, HERA and Tevatron \cite{BWHad}.
$\sqrt{s}$ is the center of mass energy of the total process;
${\mathcal L}$ is the total integrated luminosity;
 $f_0$ is the bound on the brane tension scale for one
massless branon ($N=1$) and $M_0$ is the limit on the branon mass
for small tension $f\rightarrow0$.
It is important to note that the effective approach taking in this
analysis that allows to write Lagrangian (\ref{lag}) is not valid for energy
scales $\Lambda\gtrsim 4\,\pi^{1/2}f\,N^{-1/4}$ \cite{BWRad}. Therefore, the $M_0$ value cannot be
trusted. In any case, we are providing this number since it is the most simple
 way to characterize the sensitivity of the analysis for heavy branons.
}}
\label{tabHad}
\end{table}

The most important process for branon production in a proton-proton
collider, as the LHC, is given by the gluon fusion giving a gluon
and a branon pair; as well as  a quark-gluon interaction giving a
quark and a branon pair. In these cases, the expected experimental
signal is one monojet $J$ and missing energy and momentum which can
easily be identified. An additional interesting process is the
quark-antiquark annihilation giving a photon and a branon pair. For
this process, the signature is one single photon and missing energy
and momentum. All the branons are assumed to be degenerated with a
common mass M and the quarks are considered massless. Top quark
production has been neglected due to its large mass.

The cross-section of the subprocess $g g \rightarrow g \pi\pi$ is
given by \cite{BWHad}

\begin{eqnarray}
&&\frac{d\sigma(g g \rightarrow g\pi\pi)}{dk^2dt}=\frac{\alpha_s N
(k^2-4M^2)^2}{40960f^8\pi^2\hat s^3tu}\sqrt{1-\frac{4M^2}{k^2}} \nonumber  \\
&&(\hat s^4+t^4+u^4-k^8+6k^4(\hat s^2+t^2+u^2)-4k^2(\hat
s^3+t^3+u^3)),
\end{eqnarray}
where $\hat s\equiv(p_1+p_2)^2$, $t \equiv(p_1-q)^2$,
 $u\equiv(p_2-q)^2$ and $k^2\equiv(k_1+k_2)^2$.
$p_1$ and $p_2$ are the initial gluon four-momenta, $q$ the final
gluon four-momentum and $k=k_1+k_2$ the total branon four-momentum.
Therefore the contribution to the total cross section for the $p
p\rightarrow g\pi\pi$ reaction coming from this subprocess is given
by
\begin{eqnarray}
\sigma_{gg}(p p\rightarrow g\pi\pi)= \int_{x_{min}}^1
dx\int_{y_{min}}^1 dy
g(y;\hat s) g(x;\hat s)   \nonumber\\
\int_{k^2_{min}}^{k^2_{max}}  dk^2 \int_{t_{min}}^{t_{max}}
dt\frac{d\sigma(g g\rightarrow g\pi\pi)}{dk^2dt}.
\end{eqnarray}
Here $g(x;s)$ is the gluon distribution function of the proton, $x$
and $y$ are the fractions of the protons energy carried by the
initial gluons. The different limits of the
 integrals can be written in terms of the cuts used to define
 the total cross-section. For example, in order to be able to
 detect clearly the monojet, a minimal value for its
 transverse energy $E_T$ and a pseudorapidity range given by $\eta_{min}$ and
$\eta_{max}$ must be imposed. Then the limits
 $k^2_{min}=4M^2$, $k^2_{max}=\hat s(1-2E_T/\sqrt{\hat s})$ and
 $t_{min(max)}=-(\hat s-k^2)[1+\tanh{(\eta_{min(max)})}]/2$ are obtained.
 Also we have $x_{min}=s_{min}/s$ and $y_{min}=x_{min}/x$ where
$s$ is the total center of mass energy squared of the process and
\begin{equation}
s_{min}=2E_T^2+4M^2+2E_T\sqrt{E_T^2+4M^2}.
\end{equation}

For the $q g \rightarrow q \pi\pi $ subprocess the cross-section is
given by \cite{BWHad}
\begin{eqnarray}
&&\frac{d\sigma (qg\rightarrow q\pi \pi )}{dk^{2}dt}  \nonumber \\
&=&-\frac{\alpha _{s}N}{2}\frac{(k^{2}-4M^{2})^{2}}{184320f^{8}\pi ^{2}\hat{s%
}^{3}tu}\sqrt{1-\frac{4M^{2}}{k^{2}}}(uk^{2}+4t\hat{s})(2uk^{2}+t^{2}+\hat{s}%
^{2}),
\end{eqnarray}%
with  $p_1$ and $p_2$ being the quark and the gluon four-momenta
respectively, $q$ the final state quark  four-momentum and $k_1$ and
$k_2$ the branon four-momenta. The Mandelstam variables are defined
as in previous cases.

The total cross section for the the reaction $p p\rightarrow
q\pi\pi$ is then
\begin{eqnarray}
\sigma(p p\rightarrow q\pi\pi)= \int_{x_{min}}^1 dx\int_{y_{min}}^1
dy \sum_q
 g(y;\hat s) q_p(x;\hat s)   \nonumber\\
\int_{k^2_{min}}^{k^2_{max}}  dk^2 \int_{t_{min}}^{t_{max}}
dt\frac{d\sigma(q g\rightarrow q\pi\pi)}{dk^2dt}
\end{eqnarray}

In this equation $x$ and $y$ are the fractions of  the two protons
energy carried by the subprocess quark and gluon. The different
limits of the integrals can be written as in the previous case in
terms of the the minimal transverse energy of the quark (monojet)
$E_T$.

Considering all the above equations, it is possible to compute the
total cross section $\sigma(p p\rightarrow J\pi\pi)$ in terms of the
cut in the jet transverse energy $E_T$.

On the other hand, to analyze the single photon channel, we need
the cross-section of the subprocess $q \bar q
\rightarrow \gamma \pi\pi$, that was computed in \cite{BWHad}:
\begin{eqnarray}
&&\frac{d\sigma (qq\rightarrow \gamma \pi \pi )}{dk^{2}dt}  \notag \\
&=&\frac{Q_{q}^{2}\alpha N(k^{2}-4M^{2})^{2}}{184320f^{8}\pi ^{2}\hat{s}%
^{3}tu}\sqrt{1-\frac{4M^{2}}{k^{2}}}(\hat{s}k^{2}+4tu)(2\hat{s}%
k^{2}+t^{2}+u^{2}).
\end{eqnarray}%
And this is the only leading contribution:
\begin{eqnarray}
\sigma(p p\rightarrow \gamma\pi\pi)= \int_{x_{min}}^1
dx\int_{y_{min}}^1 dy \sum_q
 \bar q_{ p}(y;\hat s) q_{ p}(x;\hat s)   \nonumber\\
\int_{k^2_{min}}^{k^2_{max}}  dk^2 \int_{t_{min}}^{t_{max}}
dt\frac{d\sigma(q q \rightarrow \gamma\pi\pi)}{dk^2dt}
\end{eqnarray}

By using all these cross sections, it is possible to compute the
expected number of events in these channels produced at the LHC in terms of the
brane tension scale $f$, the branon mass M, and the number of
branons N. The main source of uncertainty is coming from the parton
distribution function that we have taken from \cite{martin}.

Previous constraints on the branon model parameter for tree-level
processes from other colliders are summarized in Table \ref{tabHad}
\cite{BW2,Coll,BWHad}. In this Table, the present restrictions coming from
HERA, Tevatron and LEP-II are compared with the present LHC bounds
running at a center of mass energy (c.m.e.) of 7 TeV and the prospects
for the LHC running at 14 TeV c.m.e. with full luminosity. For the
single photon channel, ATLAS has published two different analysis
with a total integrated luminosity of 0.68 $nb^{-1}$ and 0.53
$nb^{-1}$ \cite{adriani}. The results are not well understood within
present hadron interaction models, but we can deduce the constraints
showed in Table \ref{tabHad} by assuming a conservative approach. In
any case, it is much more interesting the analysis presented by CMS
about jet and missing transverse energy, that is the most
constraining one for branon phenomenology due to its high
luminosity: 11.7 $nb^{-1}$ \cite{weng}. In order to complete the
analysis, we have included on Table \ref{tabHad}, the analogous
study by ATLAS which is not competitive due to the low luminosity of
0.34 $nb^{-1}$ used in the analysis \cite{schw}.

In addition, it has been shown that branon loops introduce new
couplings among standard model particle that can be described by an
effective Lagrangian. The most relevant terms of this
effective Lagrangian are \cite{BWRad}:
\begin{eqnarray}\label{eff}
{\mathcal L}_{SM}^{(1)}\simeq \frac{N \Lambda^4}{192(4\pi)^2f^8}
\left\{2T_{\mu\nu}T^{\mu\nu}+T_\mu^\mu T_\nu^\nu\right\}\,.
\end{eqnarray}
The $\Lambda$ parameter appears when dealing with branon radiative
corrections since the Lagrangian (\ref{lag}) is not renormalizable.
This parameter is the cutoff which limits the validity of the
effective description of branon and SM dynamics. A one-loop
calculation with the new effective four-fermion vertices coming from
(\ref{eff}) is equivalent to a two-loop computation with the
Lagrangian in (\ref{lag}), and it also allows to obtain the
contribution of branons to the anomalous magnetic moment of the muon
\cite{BWRad}:
\begin{equation}\label{gb}
\delta a_\mu \simeq \frac{5\, m_\mu^2}{114\,(4\pi)^4}
  \frac{N\Lambda^6}{f^8}
\end{equation}
where $N$ is the number of branon species.

The most relevant branon loops that could have compatible effects
with SM phenomenology could be the four-fermion interactions or the
fermion pair annihilation into two gauge bosons \cite{BWRad}. Bounds
on the parameter combination $f^2/(\Lambda N^{1/4})$ can be
established by considering current data. Present results are shown
in Table \ref{radcoll}, where it is also possible to find the
prospects for LHC \cite{BWRad}. However, in contrast to branon direct
production, current measurements by CMS or ATLAS detectors are not
able to improve the limits obtained with data coming from HERA
\cite{Adloff:2003jm}, Tevatron \cite{d0} and LEP
\cite{unknown:2004qh}.

The preferred parameter region for branon physics is given by
\cite{BWRad}
\begin{equation}
\text{6.0 GeV }\gtrsim \frac{f^{4}}{N^{1/2}\Lambda ^{3}}\gtrsim
\text{ 2.2 GeV (95\%c.l.)}
\end{equation}
As a result of this relation and the limits shown in Table
\ref{radcoll}, the first branon signals at colliders would be
associated to radiative corrections \cite{BWRad}.

\begin{table}[bt]
\centering
\begin{tabular}{||c|c c c||} \hline
Experiment          & $\sqrt s$ (TeV) & ${\cal L}$ (pb$^{-1}$) &
$f^2/(N^{1/4}\Lambda)$ (GeV) \\ \hline
HERA$^{\,c}$        & 0.3             &  117                   & 52                           \\
Tevatron$^{\,a,\,b}$   & 1.8        &  127                   & 69                           \\
LEP$^{\,a}$      & 0.2             &  700                   & 59                           \\
LEP$^{\,b}$      & 0.2             &  700                   & 75
\\ \hline
LHC$^{\,b}$     & 14              & $10^5$                 & 383
\\ \hline
\end{tabular}
\caption{\label{radcoll} \footnotesize{
Limits from virtual branon searches at colliders (results at the
$95\;\%$ c.l.).  The indices $^{a,b,c}$ denote the
two-photon, $e^+e^-$ and $e^+p$ ($e^-p$) channels respectively \cite{BWRad}.
Present constraints from HERA, LEP and Tevatron are compared with
the LHC prospects. The first two columns are the same as in Table
\ref{tabHad}, and the third one corresponds to the lower bound on
$f^2/(N^{1/4}\Lambda)$. In this case, current LHC constraints are
not competitive.}}
\end{table}

\section{Conclusions}\label{conclusion}

\begin{figure}[bt]
\begin{center}
\resizebox{8.8cm}{6.4cm}
{\includegraphics{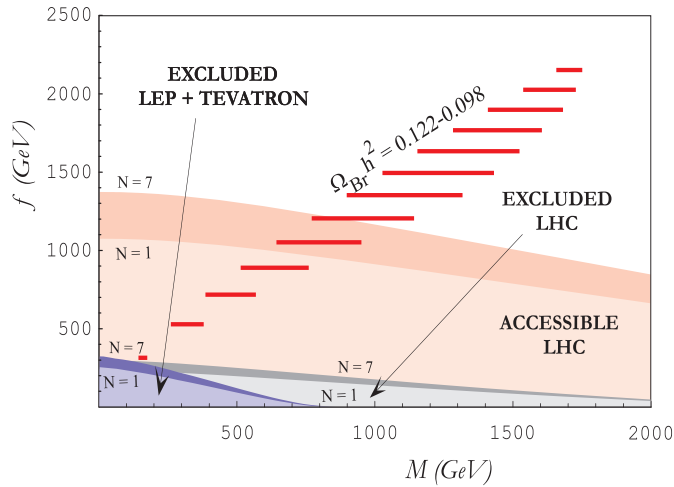}} \caption {\footnotesize The
shaded area shows the parameter space of the branon model with a
thermal relic in the range: $\Omega_{Br}h^2=0.122 - 0.098$, a
contribution to the muon anomalous magnetic moment: $a_\mu=(29\pm
18)\times 10^{-10}$ and favored by the CDMS data. The lower area is
excluded by single-photon processes at LEP together with monojet
signals at Tevatron \cite{BW2,BWHad}. Jet and missing energy analysises
perfomed with LHC data are
the most constraining for heavy branons (intermediate area).
Prospects for the sensitivity at the LHC for real branon production
are plotted also for the monojet analysis for a total integral
luminosity of ${\cal L}=10^5$ and total energy in the center of mass
of the collision of $\sqrt{s}=14$ TeV. The explicit dependence on
the number of branons $N$ is presented, since all these regions are
plotted for the extreme values $N=1$ and $N=7$. } \label{forlilian1}
\end{center}
\end{figure}

Brane fluctuations are DM candidates in brane world models with low
tension. We have studied the current situation of branons in
relation with direct dark matter detection experiments and recent
LHC measurements. As it is possible to observe in Figure
\ref{forlilian1}, the results show that, in a certain range of the
parameters for the brane tension $f$ and branon mass $M$, the relic
abundance could explain the missing mass problem and that such
parameter regions would be compatible with DM candidate events
currently observed by direct search experiments, if these events are
due to branon-nucleus coherent interactions.

Although the present low luminosity of the LHC, present constraints
from ATLAS and CMS are the most constraining for heavy branons, improving
previous analyses performed with LEP-II and TEVATRON data. In any case,
for light branons, these analises are still the most important. The same
situation is found for standard model processes mediated by virtual branon,
where current LHC measurements are not competitive yet.

In addition, there are other signatures which can prove or disprove
the model. It could be possible to detect branons in DM indirect
search experiments \cite{indirect}. Two branons may annihilate into
ordinary SM matter. Their annihilation in places like the galactic
halo, the Sun, the Earth, etc. could produce cosmic rays to be
discriminated through distinctive signatures from the background.
After annihilation, a cascade process would occur and particles such
as neutrinos, gamma rays, positrons or antimatter may be detectable
through different experiments such as Atmospheric Cerenkov
Telescopes, Neutrino Telescopes or Satellite Detectors. Work is in
progress in this direction \cite{progress}.

\section*{Acknowledgments}\label{acknowledgments}

This work is supported in part by DOE grant FG02-94ER40823, FPA
2005-02327 project (DGICYT, Spain), CAM/UCM 910309 project, MICINN
Consolider-Ingenio MULTIDARK CSD2009-00064, CONACYT (Mexico) and SNI
(Mexico).

\end{document}